\documentclass[a4paper,12pt]{article}

\usepackage[english]{babel}
\usepackage[cp1251]{inputenc}

\usepackage{amsmath,amsfonts}
\usepackage{cite}

\usepackage{lamsarrow}
\usepackage{pb-diagram,pb-lams}

\def\Z{\mathbb{Z}}
\def\res {\operatorname{res}}
\def\dfrac#1#2{\frac{\partial #1}{\partial #2}}
\title{On the integrability of a lattice equation  \\ with two continuum limits}

\author{R.N. Garifullin and R.I. Yamilov}
\begin{document}







\maketitle
\begin{abstract} 
We study a new example of lattice equation being one of the key equations of a recent generalized symmetry classification of five-point differential-difference equations. This equation has two different continuum limits which are the well-known fifth order partial-differential equations, namely, the Sawada-Kotera and Kaup-Kupershmidt equations. We justify its integrability by constructing an $L-A$ pair and a hierarchy of conservation laws. 
\end{abstract}



\maketitle


\section{Introduction}

We consider the differential-difference equation
\begin{equation}{u_{n,t} = (u_n+1)\left(\frac{u_{n+2}u_n(u_{n+1}+1)^2}{u_{n+1}}-\frac{u_{n-2}u_n(u_{n-1}+1)^2}{u_{n-1}}+(2u_n+1)(u_{n+1}-u_{n-1})\right),\label{rat}}\end{equation}
where $n\in\Z,$ while $u_n(t)$ is an unknown function of one discrete variable $n$ and one continuous variable $t$, and the index $t$ in $u_{n,t}$ denotes the time derivative. Equation \eqref{rat} is obtained as a result of the generalized symmetry classification of five-point differential-difference equations \begin{equation}{\label{gen5}u_{n,t}=F(u_{n+2},u_{n+1},u_n,u_{n-1},u_{n-2}),}\end{equation} carried out in \cite{gyl16_1,gyl17,gyl17_1}. Equation \eqref{rat} coincides with the equation \cite[(E16)]{gyl17}. Even earlier this equation has been  obtained in \cite{a16_1}. Equations of the form \eqref{gen5} play an important role in the study of four-point discrete equations on the square lattice, which are very relevant for today, see e.g. \cite{gy12,a11,mx13,gmy14}. 

At the present time there is very little information on equation \eqref{rat}. 
It has been shown in \cite{gyl17} that  equation \eqref{rat} possesses a nine-point generalized symmetry of the form \begin{equation*}u_{n,\theta}=G(u_{n+4},u_{n+3},\ldots,u_{n-4}).\end{equation*} As for relations to the other known integrable equations of the form \eqref{gen5}, nothing useful from the viewpoint of constructing solutions is known, see details in the next Section.

However, this equation occupies a special place in the classification \cite{gyl16_1,gyl17,gyl17_1}. In particular, it possesses a remarkable property discovered in \cite{gy17}. This equation has two different continuum limits being the well-known  Kaup-Kupershmidt and Sawada-Kotera equations, details will be given below. For this reason, equation \eqref{rat} deserves a more detailed study. 

In Section \ref{s1} we discuss the known properties of equation \eqref{rat}. In order to justify the integrability of \eqref{rat}, we construct an $L-A$ pair in Section \ref{s2} and show that it provides an infinity hierarchy of conservation laws in Section \ref{s3}.

\section{Special place of equation \eqref{rat} in the classification \cite{gyl16_1,gyl17,gyl17_1}}\label{s1}

In two lists of integrable equations of the form \eqref{gen5} presented in \cite{gyl17,gyl17_1}, the following four equations occupy a special place: those are \eqref{rat} and 
\begin{equation}{u_{n,t}=(u_n^2-1)(u_{n+2}\sqrt{u_{n+1}^2-1}-u_{n-2}\sqrt{u_{n-1}^2-1}),\label{d_eq}}\end{equation} 
\begin{equation}{ u_{n,t} = u_n^2(u_{n+2}u_{n+1}-u_{n-1}u_{n-2})-u_n(u_{n+1}-u_{n-1}),\label{seva}}\end{equation}
\begin{equation}{ u_{n,t} = u_{n+1}u_n^3u_{n-1}(u_{n+2}u_{n+1}-u_{n-1}u_{n-2})-u_n^2(u_{n+1}-u_{n-1}).\label{to_seva}}\end{equation}
Equations (\ref{d_eq}-\ref{to_seva}) correspond to equations (E17), (E15) of \cite{gyl17} and (E14) of \cite{gyl17_1}, respectively. Equation \eqref{seva} is known for a long time \cite{th96}.

All other equations of \cite{gyl17,gyl17_1} go over in the continuum limit into the third order equations of the form
\begin{equation}{U_{\tau}=U_{xxx}+F(U_{xx},U_{x},U),\label{Utau3}}\end{equation} where the indices $\tau$ and $x$ denote $\tau$ and $x$ partial derivatives,
and mainly into the Korteweg-de Vries equation.
These four equations correspond in the continuum limit to the fifth order equations of the form:
\begin{equation}{U_{\tau}=U_{xxxxx}+F(U_{xxxx},U_{xxx},U_{xx},U_{x},U).\label{Utau}}\end{equation}
For all the four equations (\ref{rat},\ref{d_eq}-\ref{to_seva}) we get in the continuum limit one of the two well-known equations. 
One of them is the Kaup-Kupershmidt equation \cite{fg80,k80}:
\begin{equation}{U_\tau=U_{xxxxx}+5UU_{xxx}+\frac{25}2U_xU_{xx}+5U^2U_x,\label{kk}}\end{equation} and the second one is the Sawada-Kotera equation \cite{sk74}:
\begin{equation}{U_\tau=U_{xxxxx}+5UU_{xxx}+5U_xU_{xx}+5U^2U_x\label{sk}.}\end{equation}

Using the substitution
\begin{equation}{\label{d_to_s}u_n(t)=\frac{\sqrt 3}3+\frac{\sqrt 3}{2}\varepsilon^2 U\left(\tau-\frac{18}{5}\varepsilon^5 t,x+\frac 43\varepsilon t\right),\quad x=\varepsilon n,}\end{equation} in equation \eqref{to_seva}, we get at $\varepsilon\to 0$ the Sawada-Kotera equation \eqref{sk}. All other continuum limits are known, see \cite{a11} for \eqref{seva} and \cite{gy17} for \eqref{rat} and \eqref{d_eq}.
Here we explicitly replicate substitutions for equation \eqref{rat} under study which has two different continuum limits. 
The substitution 
\begin{equation}{\label{r_kk}u_n(t)=-\frac{4}3-\varepsilon^2 U\left(\tau-\frac{18}{5}\varepsilon^5 t,x+\frac 43\varepsilon t\right),\quad x=\varepsilon n,}\end{equation} in \eqref{rat} leads to equation \eqref{kk}, while
the substitution 
\begin{equation}{\label{r_sk}u_n(t)=-\frac{2}3+\varepsilon^2 U\left(\tau-\frac{18}{5}\varepsilon^5 t,x+\frac 43\varepsilon t\right),\quad x=\varepsilon n,}\end{equation} leads  to equation \eqref{sk}. 
The link between these discrete and continuous equations is shown in the following diagram:
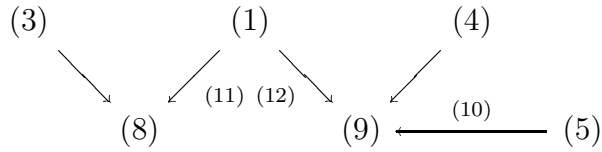
\[ \begin{diagram}
\node{\eqref{d_eq}}
\arrow{se,r}{} 
\node[2]{\eqref{rat}}
\arrow{sw,r}{\eqref{r_kk}} 
\arrow{se,r}{\eqref{r_sk}} 
\node[2]{\eqref{seva}}
\arrow{sw,r}{} 
 \\
\node[2]{\eqref{kk}}
\node[2]{\eqref{sk}}
\node[2]{\eqref{to_seva}}\arrow[2]{w,l}{\eqref{d_to_s}} 
\end{diagram}\]
We see that equation \eqref{kk} has two different integrable approximations, while equation \eqref{sk} has three approximations.

As far as we know, there are no relations between (\ref{rat},\ref{d_eq}-\ref{to_seva}) and other known equations of the form \eqref{gen5} presented in \cite{gyl17,gyl17_1}. More precisely, we mean relations in the form of the transformations \begin{equation}{\label{transf}\hat u_n=\varphi(u_{n+k},u_{n+k-1},\ldots,u_{n+m}),\ \ k>m,}\end{equation} and their compositions, see a detailed discussion of such transformations in \cite{gyl16_1}. As for relations among (\ref{rat},\ref{d_eq}-\ref{to_seva}), equation \eqref{to_seva} is transformed into \eqref{seva} by $\hat u_n=u_{n+1}u_n$, i.e. \eqref{to_seva} is a simple modification of \eqref{seva}.
There is a complicated relation between equations \eqref{rat} and \eqref{seva} found in \cite{a16_1}. As it is shown in \cite{gyl17}, it is a composition of two Miura type transformations. It is very difficult to use that relation
for the construction of solutions because the problem is reduced to solving the discrete Riccati type equations \cite{gyl17}.

There is a complete list of integrable equations of the form \eqref{Utau}, see \cite{ms12,dss85,mss91}. Equations \eqref{kk} and \eqref{sk} play the key role in that list, since all the other are transformed into these two by transformations of the form:
\begin{equation*}\hat U=\Phi(U,U_x,U_{xx},\ldots,U_{x\ldots x}).\end{equation*}

\section{$L-A$ pair }\label{s2}

As the continuum limit shows, equation \eqref{rat} should be close to equations (\ref{d_eq},\ref{seva}) in its integrability properties, and these equations (\ref{d_eq},\ref{seva}) have the $L-A$ pairs defined by $3 \times 3$ matrices \cite{a11,gy17}. Here we construct an $L-A$ pair for equation \eqref{rat} following \cite{gy17}. 

We look for an $L-A$ pair of the form
\begin{equation}{L_n\psi_n=0,\quad \psi_{n,t}=A_n\psi_n\label{LA}}\end{equation} with the operator $L_n$ of the form 
\begin{equation}{\label{form_L}L_n=T^2+l_n^{(1)}T+l_n^{(0)}+l_n^{(-1)}T^{-1},}\end{equation} where $l_n^{(k)}$ with $k=-1,0,1$ depend on a finite number of the functions $u_{n+j}.$ Here $T$ is the shift operator:
$Th_n=h_{n+1}.$ In  the case of \eqref{form_L} the operator $A_n$ can be chosen as:
\begin{equation*}A_n=a_n^{(1)}T+a_n^{(0)}+a_n^{(-1)}T^{-1}.\end{equation*}
The compatibility condition for system \eqref{LA} has the form
\begin{equation}{\label{L_t}\frac{ d(L_n\psi_n)}{dt}=(L_{n,t}+L_nA_n)\psi_n=0}\end{equation} and it must be satisfied in virtue of the equations \eqref{rat} and $L_n\psi_n=0$. 

If we suppose that the coefficients $l_n^{(k)}$ depend on $u_n$ only, then we can check that $a_n^{(k)}$ have to depend on $u_{n-1},u_n$ only. However, in this case the problem has no solution for equation \eqref{rat}. Therefore we proceed to the case when the functions $l_{n}^{(k)}$ depend on $u_n,u_{n+1}$. Then the coefficients $a_n^{(k)}$ must depend on $u_{n-1},u_n,u_{n+1}$. In this case we have managed to find operators $L_n$ and $A_n$ with one irremovable arbitrary constant $\lambda$ playing the role of the spectral parameter here:
\begin{equation}{L_n=T^2-\frac{U_{n+1}}{u_{n+1}} T+\lambda\frac{U_{n+1}}{u_n}\left(1-\frac{u_{n}}{U_n}T^{-1}\right),\label{Lsc}}\end{equation}
\begin{equation}{A_n=\frac{u_n}{U_n}(\lambda T^{-1}-\lambda^{-1}T)+\frac{u_n}{U_n^2}(u_{n-1}+u_{n+1}T^{-1})(T-1),\label{Asc}}\end{equation}
where \begin{equation}{U_n=\frac{u_n}{1+u_n}\label{U}.}\end{equation}

The $L-A$ pair (\ref{LA},\ref{Lsc},\ref{Asc}) can be rewritten in the standard matrix form with $3\times 3$ matrices $\tilde L_n, \tilde A_n$:
\begin{equation*}\Psi_{n+1}=\tilde L_n \Psi_n,\quad \Psi_{n,t}=\tilde A_n \Psi_n,\end{equation*} where $\Psi_n$ is a spectral vector-function, whose standard form is \begin{equation*}\Psi_n=\left(\begin{array}{c}\psi_{n+1}\\ \psi_n\\ \psi_{n-1}\end{array}\right).\end{equation*}
Here we slightly change $\Psi$ by gauge transformation to simplify the matrices $\tilde L_n, \tilde A_n$:
\begin{equation*}\Psi_n=\left(\begin{array}{c}U_n(\lambda\psi_{n+1}+\frac{1}{u_n}\psi_n)\\ \psi_n\\ \psi_{n-1}\end{array}\right),\end{equation*}
and now $\tilde{L}_n, \tilde{A}_n$ read:
\begin{equation}{\label{Lm}\tilde{L}_n=\left(\begin{array}{ccc}0&-\frac{1}{u_{n}}&\frac{1}{U_n}\\ \lambda U_n&\frac{U_n}{u_n} &0\\0 &1&0\end{array}\right),}\end{equation}
\begin{equation}{\label{Am}\tilde{A}_n=\left(\begin{array}{ccc}
\frac{u_{n-1}u_{n-2}}{U_{n-1}^2}-\frac{u_{n+1}u_n}{U_n^2}-u_n+u_{n-1}&-\frac{u_{n-1}}{\lambda U_{n-1}}-\frac{u_{n-1}u_{n-2}}{\lambda U_{n-1}^2}&\frac{u_{n+1}u_n+u_{n+1}+u_n}{U_{n}}\\ \lambda(1+u_n)u_{n-1}-u_n&\frac{u_{n+1}u_n+u_{n+1}-u_nu_{n-1}}{U_n}-\frac{1}{\lambda} &\frac{\lambda u_{n}}{U_n}-\frac{u_{n+1}u_n}{U_n^2}\\
\lambda u_{n-1}-(1+u_{n-1})u_n &\frac{u_{n-1}u_{n-2}}{U_{n-1}^2}-\frac{u_{n-1}}{\lambda U_{n-1}}&\lambda-\frac{u_{n-1}u_{n-2}}{U_{n-1}^2}+\frac{u_nu_{n-1}}{U_{n-1}}\end{array}\right).}\end{equation}
In this case, unlike \eqref{L_t}, the compatibility condition can be represented in a form which does not use the spectral vector-function $\Psi_n$. It will be the following matrix form in terms of $3\times 3$ matrices:
\begin{equation}{\tilde L_{n,t}=\tilde A_{n+1}\tilde L_n-\tilde L_n \tilde A_n\label{LAmat}.}\end{equation}

\section{Conservation laws}\label{s3}

As far as we know, there exist two methods to construct the conservation laws by using the matrix $L-A$ pair (\ref{LAmat}), see \cite{hy13,gmy14,m15}. However, we do not see how to apply those methods in the case of matrices \eqref{Lm} and \eqref{Am}. Here we will use a different scheme, presented in \cite{gy17}, for deriving conservation laws from the $L-A$ pair \eqref{LA}. In \cite{gy17} that scheme was applied to one equation \eqref{d_eq} only. Here we check  it again by example of one more equation \eqref{rat}.

The structure of operators (\ref{Lsc},\ref{Asc}) allows us to rewrite the $L-A$ pair \eqref{LA} in form of the Lax pair. The operator $L_n$ has a linear dependence on $\lambda$:
\begin{equation}{L_n=P_n-\lambda Q_n,\label{Lpq}}\end{equation}
where
\begin{equation*}P_n=T^2-\frac{U_{n+1}}{u_{n+1}} T,\quad Q_n=-\frac{U_{n+1}}{u_n}\left(1-\frac{u_{n}}{U_n}T^{-1}\right),\end{equation*} and $U_n$ is defined by \eqref{U}.
Introducing $\hat L_n=Q_n^{-1}P_n$ we get an equation of the form:
\begin{equation}{\hat L_n \psi_n=\lambda \psi_n.\label{hL}}\end{equation}
The functions $\lambda \psi_n$ and $\lambda^{-1}\psi_n$ in the second equation of \eqref{LA} can be expressed in terms of $\hat L_n$ and $\psi_n$, using \eqref{hL} and its consequence 
$\lambda^{-1} \psi_n=\hat L_n^{-1} \psi_n.$ As a result we have:
\begin{equation}{\psi_{n,t}=\hat A_n \psi_n,\label{hA}}\end{equation} where
\begin{equation*}\hat A_n=\frac{u_n}{U_n}( T^{-1}Q_n^{-1}P_n-TP_n^{-1}Q_n)+\frac{u_n}{U_n^2}(u_{n-1}+u_{n+1}T^{-1})(T-1).\end{equation*}

It is important that the new operators $\hat L_n$ and $\hat A_n$ in the $L-A$ pair (\ref{hL},\ref{hA}) do not depend on the spectral parameter $\lambda$. For this reason, the compatibility condition can be written in the operator form, without using the $\psi$-function:
\begin{equation}{\hat L_{n,t}=\hat A_{n}\hat L_n-\hat L_n \hat A_n=[\hat A_n,\hat L_n],\label{comp_h}}\end{equation} and this is nothing but the Lax equation.
The difference between this $L-A$ pair and the well-known Lax pairs for the Toda and Volterra equations is that the operators $\hat L_n$ and $\hat A_n$ are nonlocal. Nevertheless, using the definition of inverse operators: \begin{equation}{P_n P_n^{-1}=P_n^{-1}P_n=1, \quad Q_nQ_n^{-1}=Q_n^{-1}Q_n=1\label{invop}}\end{equation}and the fact that they are linear, we can check that \eqref{comp_h} is true by direct calculation.

The conservation laws of equation \eqref{rat}, which are expressions of the form 
\begin{equation*}\rho_{n,t}^{(k)}=(T-1)\sigma_n^{(k)},\ k\geq 0,\end{equation*}
can be derived from the Lax equation \eqref{comp_h}, notwithstanding the nonlocal structure of $\hat L_n,\hat A_n$, see \cite{y06}. For this, first of all, we have to represent the operators $\hat L_n,\ \hat A_n$ as formal series in powers of $T^{-1}$:
\begin{equation}{H_n=\sum_{k\leq N}h_n^{(k)}T^k.\label{opH}}\end{equation} 
Formal series of this kind can be multiplied according the rule: $(a_nT^k)(b_nT^j)=a_nb_{n+k}T^{k+j}.$ The inverse series of the form \eqref{opH} can be easily obtained by definition \eqref{invop}, for instance: 
\begin{equation*}Q_n^{-1}=-(1+q_nT^{-1}+(q_nT^{-1})^2+\ldots+(q_nT^{-1})^k+\ldots)\frac {u_n}{U_{n+1}},\ \ q_n=\frac{u_{n}}{U_n}.\end{equation*}
The series $\hat L_n$ has the second order:
\begin{equation*}\hat L_n=\sum_{k\leq 2}l_n^{(k)}T^k=-\left(\frac{u_n}{U_{n+1}}T^2+u_n\left(\frac{u_{n-1}}{U_n^2}-\frac{1}{u_{n+1}}\right)T+\frac{u_{n-1}}{U_{n}}\left(\frac{u_{n}u_{n-2}}{U_{n-1}^2}-1\right)T^0+\ldots\right).\end{equation*}

The conserved densities $\rho_{n}^{(k)}$ of equation \eqref{rat} can be found as: \begin{equation}{\rho_n^{(0)}=\log l_n^{(2)},\qquad \rho_n^{(k)}=\res \hat L_{n}^k,\ \  k\geq 1, \label{form_d}}\end{equation} where the residue of  formal series \eqref{opH} is defined by the rule: $\res H_n=h_n^{(0)}$, see \cite{y06}. Corresponding functions $\sigma_n^{(k)}$ can easily be found by direct calculation.

The conserved densities $\hat \rho_n^{(k)}$ below have been found in this way and then simplified in accordance with the rule:
\begin{equation*}\hat\rho_n^{(k)}=c_k\rho_n^{(k)}+(T-1)g_n^{(k)}, \end{equation*} where $c_k$ is a constant and $g_n^{(k)}$ is a function. The first three densities of equation \eqref{rat} read:
\begin{equation*}\hat\rho_n^{(0)}=\log(u_n+1),\end{equation*}
\begin{equation*}\hat\rho_n^{(1)}=\frac{V_{n+1}u_{n-1}}{U_n},\end{equation*}
\begin{align*}\hat\rho_n^{(2)}&=\frac{u_{n+2}u_{n+1}u_n^2u_{n-1}u_{n-2}}{U_{n+1}^2U_nU_{n-1}^2}+\frac{u_{n+1}u_{n-2}(V_n^2-u_{n}u_{n-1})}{U_{n}U_{n-1}}+\frac{u_{n+1}^2u_{n}^2u_{n-1}}{2U_{n+1}^2U_{n}^3}\\&-\frac{u_{n+1}u_{n}u_{n-1}}{U_{n+1}U_{n}}+\frac{u_{n+1}u_{n-1}(V_{n+1}-1)V_n}{2U_{n+1}U_{n}^2}+\frac{u_{n-1}^2}{2U_n^2}\nonumber,\end{align*}
where \begin{equation*}V_{n}=u_{n}u_{n-1}+u_{n}+u_{n-1}.\end{equation*}

We can easily check that
\begin{equation*}\dfrac{^2\hat\rho_n^{(1)}}{u_{n+1}\partial u_{n-1}}\neq0,\qquad\dfrac{^2\hat\rho_n^{(2)}}{u_{n+2}\partial u_{n-2}}\neq0.\end{equation*}
Therefore, in accordance with a theory of the review \cite{y06}, the conserved densities $\hat\rho_n^{(0)},\ \hat\rho_n^{(1)},\ \hat\rho_n^{(2)}$ have the orders $0,2,4$ respectively. This means that we have got three conserved densities, which are nontrivial and essentially different.



\end{document}